\documentclass[useAMS,usenatbib,usegraphicx]{mn2e}
\usepackage{aas_macros,times}

\title[Red-Sequence Dwarf-to-Giant Ratios]{Tracking Down a Critical Halo Mass for Killing Galaxies through the Growth of the Red-Sequence}
\author[Gilbank \& Balogh.]{David G.\ Gilbank\thanks{Email: dgilbank@astro.uwaterloo.ca} and
  Michael L.\ Balogh\thanks{Email: mbalogh@uwaterloo.ca}\\
  Astrophysics and Gravitation Group, Department of Physics \& Astronomy, University of Waterloo, Waterloo, Ontario, Canada N2L 3G1}

\pagerange{\pageref{firstpage}--\pageref{lastpage}} \pubyear{2007}

\def\LaTeX{L\kern-.36em\raise.3ex\hbox{a}\kern-.15em
    T\kern-.1667em\lower.7ex\hbox{E}\kern-.125emX}
\def\gsim{\mathrel{\raise0.35ex\hbox{$\scriptstyle >$}\kern-0.6em
\lower0.40ex\hbox{{$\scriptstyle \sim$}}}}
\def\lsim{\mathrel{\raise0.35ex\hbox{$\scriptstyle <$}\kern-0.6em
\lower0.40ex\hbox{{$\scriptstyle \sim$}}}}

\begin{document}

\label{firstpage}

\maketitle

\begin{abstract}
Red-sequence galaxies record the history of terminated star-formation in the Universe and can thus provide important clues to the mechanisms responsible for this termination.  We construct composite samples of published cluster and field galaxy photometry in order to study the build-up of galaxies on the red-sequence, as parameterised by the dwarf-to-giant ratio (DGR).  We find that the DGR in clusters is higher than that of the field at all redshifts, implying that the faint end of the red-sequence was established first in clusters.  We find that the DGR evolves with redshift for both samples, consistent with the ``down-sizing'' picture of star formation.  We examine the predictions of semi-analytic models for the DGR and find that neither the magnitude of its environmental dependence nor its evolution is correctly predicted in the models.  Red-sequence DGRs are consistently too high in the models, the most likely explanation being that the strangulation mechanism used to remove hot gas from satellite galaxies is too efficient.  Finally we present a simple toy model including a threshold mass, below which galaxies are not strangled, and show that this can predict the observed evolution of the field DGR.
\end{abstract}


\section{Introduction}
Red-sequence galaxies are an important population for understanding galaxy formation and evolution in general.  They represent systems with very little or no on-going star-formation and thus are a unique tracer of the past activity of galaxies.  \citet{De-Lucia:2004xa} found a deficit of faint red-sequence galaxies in clusters at z$\sim$0.8 relative to local clusters.  This would suggest that star formation ended earlier for the most luminous/massive galaxies at high redshift and would support the ``down-sizing'' picture first proposed by \citet{Cowie:1996xw}, in which the termination of star-formation progresses from the most massive to the least massive galaxies as the universe ages.  Similar results for clusters spanning a range of redshifts have been found by \citet{Tanaka:2005mk}, \citet{de-Lucia:2007li}, \citet{Stott:2007wc}, \citet{Gilbank:2007rq} and others.  Similarly, surveys of field galaxies have attempted to characterise the evolution of the red-sequence luminosity function.  The relative contributions of passive evolution/termination of star-formation and number density evolution/dry merging are still open questions (e.g., \citealt{Bell:2004lb},  \citealt{Cimatti:2006vz}).

Red-sequence galaxies can thus place constraints on galaxy formation prescriptions in theoretical models.  Recent observations in the local universe using SDSS data \citep{Baldry:2006bq,Weinmann:2006cn} have found that the fractions of satellite galaxies which are red are too high in current semi-analytic models.  With the recent abundance of surveys targeting red-sequence galaxies both in the cluster and the field, the time is now ripe to draw together these works to build a picture of the build-up of galaxies on the red-sequence as a function of environment and to confront these with predictions of galaxy formation models.

\section{Observational Data}
We aim to construct the most uniform sample possible using colour-selected red-sequence galaxy data from the literature.  To this end, we transform all our data onto the same system.  A simpler quantity than the luminosity function (LF), which has been employed in the past, is the red-sequence dwarf-to-giant ratio, hereafter DGR.  This is essentially just a LF reduced to two bins and avoids the complications of having to fit an analytic function (usually with degeneracies between the fitted parameters) to the data.
  Following \citet{De-Lucia:2004xa}, we define luminous, or giant, galaxies as those brighter than $M_{\rm V}=-20.0$ and faint, or dwarfs, as those with $-20<M_{\rm V}\le-18.2$ (k- and evolution-corrected). Where data are presented on a different photometric system, we use \citet{Bruzual:2003de} stellar population models to calculate the transformation using a $z_f=3$, Solar metallicity, single stellar population with a Chabrier IMF. Where DGRs have been given in the original paper using these limits (as is the case for much of the cluster data) we use these numbers and their associated errors directly.  For other works, we take the $M^\star$ and $\alpha$ values from the Schechter function fit, transform to $M_{\rm V}$ and integrate the transformed LF to calculate DGR.  We estimate conservative errors using a Monte Carlo method taking the errors in the fitted parameters and allowing for their covariance by employing the error ellipse given in the original reference.  In addition we propagate a 0.05 magnitude uncertainty (discussed below) in the magnitude limits to allow for any residual mismatch in the converted photometric system.   

For the field samples, we carefully examine all of the Schechter function fits in each redshift bin and only use measurements where the data fully cover the dwarf and giant range.  For example, the \citet[][NOAODWFS]{Brown:2007kk} data do a good job of constraining the shape of the LF over the magnitude range we require for z$\le$0.5 but at higher redshift are insufficiently deep to constrain the number of dwarfs.  Conversely, the \citet[][VVDS]{Zucca:2006oz} data are sufficiently deep to measure the faint end of the LF at higher redshift, but at z$\le$0.6 are subject to considerable uncertainty around $M^\star$, presumably due to the relatively limited area of the survey.  The remainder of our field sample uses points from \citet[][SDSS]{Baldry:2004wj}, \citet[][SDSS]{Bell:2003iq}, \citet[][COMBO-17]{Bell:2004lb}, \citet[][Millennium Galaxy Catalogue]{Driver:2006li} and \citet[][COSMOS]{Scarlata:2007oz}.  Fig.~1 shows DGR as a function of redshift for these surveys.  Where the redshift ranges of different surveys overlap, we find good agreement in our measured DGRs (e.g., VVDS and COSMOS at $z=0.7$).

\begin{figure*}
{\centering
\includegraphics[width=140mm]{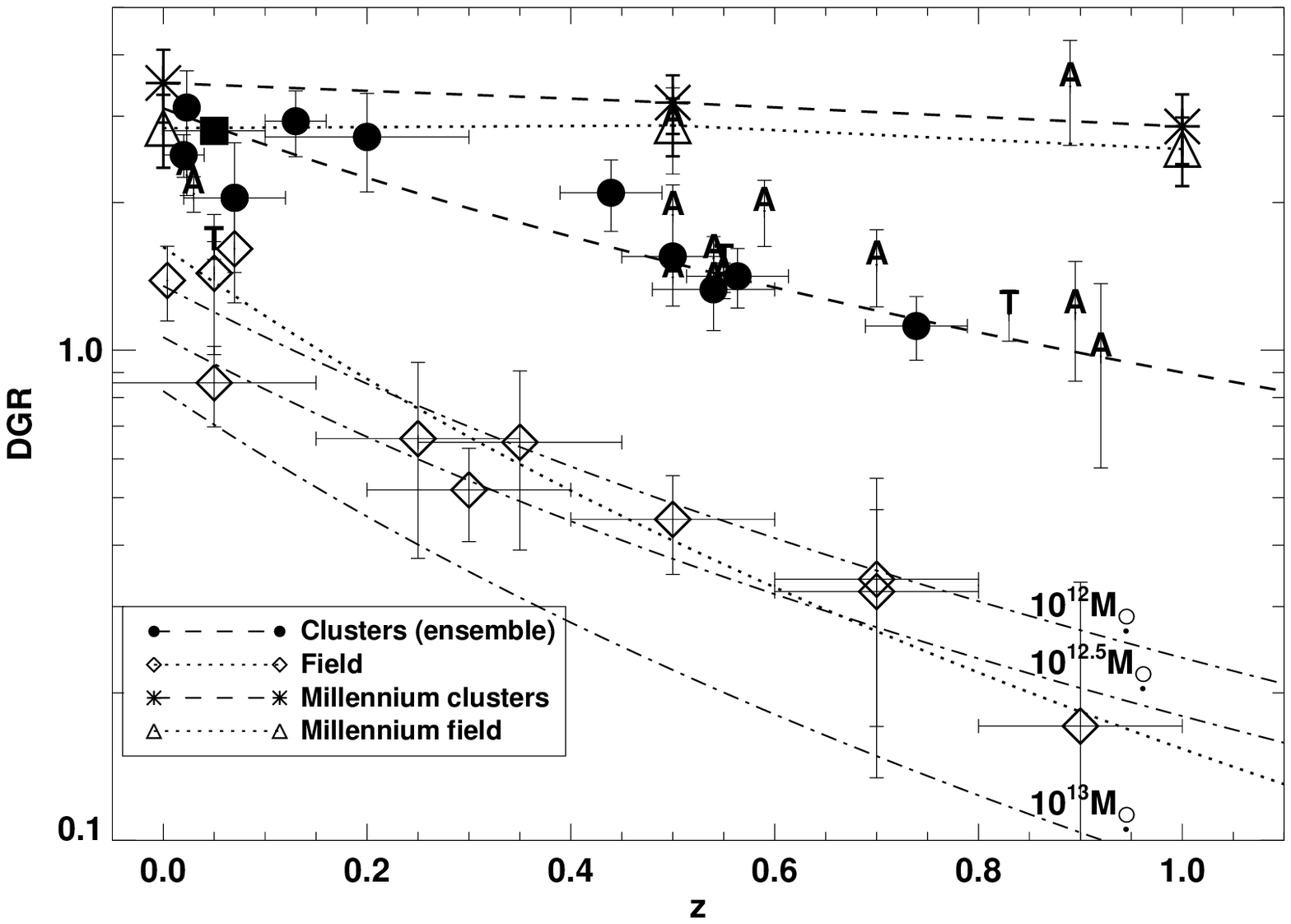}
\caption{The evolution of the Dwarf-to-Giant ratio, DGR, for various samples taken from the literature and converted to a uniform system, as described in the text.  Filled circles and open diamonds denote cluster (\citealt{barkhouse07a}, \citealt{de-Lucia:2007li}, \citealt{Gilbank:2007rq}, \citealt{Hansen:2007bb}, \citealt{Stott:2007wc}) and field (\citealt{Bell:2003iq}, \citealt{Baldry:2004wj}, \citealt{Bell:2004lb}, \citealt{Driver:2006li}, \citealt{Zucca:2006oz}, \citealt{Brown:2007kk}, \citealt{Scarlata:2007oz}) samples respectively from observational data, and lines show $(1+z)^\beta$ fits.  Horizontal error bars indicate the redshift range covered by the data.  Naked error bars are measurements for individual clusters from Andreon (2007) and \citet{Tanaka:2005mk}, labelled `A' and `T' respectively.  Only ensemble clusters are included in the fit.  The filled square shows the highest density regions in the SDSS from \citet{Baldry:2006bq} which agrees well with the cluster data.  Asterisks and triangles show predictions from the \citet{Bower:2006vw} semi-analytic model, connected with lines to guide the eye.  The dot-dashed lines show the predicted evolution of the field DGR if isolated galaxies have a constant DGR=0 and the observed trend is due to the increasing abundance of groups  (having the same DGR as clusters) above the labelled threshold mass, $M_{th}$ with cosmic time, as described in \S4.
\label{fig:dgrz}
}
}
\end{figure*}

Clusters show considerable scatter in their individual DGRs (e.g., \citealt{de-Lucia:2007li}) and so we only focus on works measuring DGR from ensemble averages of clusters in each redshift bin (\citealt{de-Lucia:2007li}, \citealt{Stott:2007wc}, \citealt{Gilbank:2007rq}, \citealt{Hansen:2007bb}, \citealt{barkhouse07a}). 

Two results are immediately apparent from the observational data in Fig.~1: the DGR of field red-sequence galaxies is always lower than that of cluster galaxies at the same redshift, and the DGR of both samples evolves with redshift.  We parameterise the evolution of the DGR with a fit of the form $(1+z)^\beta$ finding $\beta=-1.8\pm0.5$ and $\beta=-3.4\pm1.1$ for the cluster and field samples respectively.

\subsection{Sources of systematic error}

The first systematic to be addressed is that of transforming published photometry on different systems to a uniform one.  Most of the published cluster studies give rest-frame $V$ magnitudes (e.g., \citealt{de-Lucia:2007li}), and most field studies give rest-frame $B$ (e.g., \citealt{Bell:2004lb}).  This is not exclusively the case, however, and we find good agreement where different studies measured in different passbands in the same environments at similar redshifts overlap \citep[e.g.,][]{Hansen:2007bb,Stott:2007wc}.  Thus we believe the transformation is not a large source of error.  We have propagated a generous 0.05 magnitude error through our analysis to allow for this. This also allows for the dominant transformation uncertainty which is that of the systematic calibration in the individual surveys (e.g., Hansen et al. 2007 quote their calibration uncertainty as 0.03 mag). In fact, comparing  \citet{Hansen:2007bb} and \citet{Stott:2007wc}, the measured DGRs agree within the errors even if this additional uncertainty is neglected.  So, we believe that our error bars are ample.

The important other systematic for field studies is cosmic variance.  The authors of the various works included here discuss the size of this effect for their surveys, and this is not likely to be a major contribution to the error bars in Fig.~1.  Cluster studies require more care in the sample selection to minimise systematic errors.  As found by e.g., \citet{Gilbank:2007rq} and \citet{Hansen:2007bb}, the faint end slope of the cluster RSLF depends on cluster richness/mass (at least for z$\lsim$0.5), in the sense that low mass clusters/rich groups are expected to have systematically lower DGRs.  In this compilation, we have attempted to utilise the richest/most massive systems at each redshift for the cluster sample, to maximise the difference between the clusters and the field.  In so much as, on average,  the most massive systems at one redshift evolve into the most massive systems at lower redshift, such a selection is appropriate for evolutionary studies.  If we consider the three data points nearest to z$\sim$0.5, these were drawn from very differently selected samples: a statistical sample of 54 red-sequence selected clusters, having richnesses corresponding to a mass limit of $\gsim 6 \times 10 ^{13} $M$_\odot$ 
\citep{Gilbank:2007rq}; a sample of 11 clusters selected via their unresolved cluster light from shallow optical imaging, having various masses \citep{de-Lucia:2007li}; and a sample of ten of the most X-ray luminous clusters \citep{Stott:2007wc}.  The good agreement of the points from these ensemble clusters suggests that the DGR is relatively insensitive to mass or selection method over this mass range.  Furthermore, it suggests that the evolution of the cluster DGR is likely not due to a systematically changing mass limit with redshift.

Additionally, the virial radius of a cluster depends on its mass roughly as $r_{vir} \propto M^{1/3}$ \citep[e.g.,][]{carlberg97}, so using a fixed physical radius within which to measure a property of a cluster probes a different fraction of the cluster's extent depending on the mass of the cluster.  The effect of this on DGR is small as can be seen from fig.~5 of \citet{de-Lucia:2007li}.  By limiting the mass range of the clusters to the most massive systems, we have minimised the possible impact of such an effect, since the difference between a fixed physical radius and a fixed fraction of the virial radius only depends on the mass to the $1/3$ power.

\subsection{Conflicting results}

 \citet{Andreon:2007jb} recently argued, based on a small number of individual clusters, that the cluster DGR does not appear to evolve with redshift.  We indicate his data points on Fig.~1, omitting the three z$>$1 clusters for which the data do not bracket the 4000\AA~break.  All of the individual clusters are in reasonable agreement with the  ensemble averages, and with our best fit line, except for one at $z=0.89$.  Furthermore, his sample only contains two $z<0.5$ clusters, one of which is Coma, for which the DGR is in good agreement with the \citet{de-Lucia:2007li} value.  Given the significant cluster-to-cluster scatter, the ensemble samples compiled here provide a better reflection of the global average in clusters, and allow us to measure significant evolution with redshift.

\citet{Tanaka:2005mk} and \citet{de-Lucia:2007li} both constructed composite samples of local SDSS clusters from the C4 cluster catalogue \citep{Miller:2005qb}.  However,  \citet{Tanaka:2005mk} found a low DGR, comparable with their two higher redshift cluster z$=$0.55 and z$=$0.83) and therefore concluded that the cluster DGR might not evolve with redshift.  Both of their higher-z individual clusters lie on the best-fit relation of Fig.~1, so the discrepancy arises from the z$\sim$0 point.  \citet{de-Lucia:2007li} found a higher DGR (for their $\sigma >$ 600 km s$^{-1}$ sample) more consistent with the results of \citet{Hansen:2007bb} for the richest clusters; \citet{Stott:2007wc} for the most X-ray luminous clusters; and also for the regions of highest density in the SDSS \citep{Baldry:2006bq} (open square).  We suggest that the reason for the low DGR measured by \citet{Tanaka:2005mk} may be that their sample was dominated by lower mass systems.  We note that their measurement is just consistent with the lowest DGR point at z$\sim$0.1 in Fig.~1 which comes from \citet{barkhouse07a} and comprised a set of 57 Abell clusters over a range of richnesses.

\section{Predictions from the Galform semi-analytic models}

We draw samples from the Millennium simulation using the semi-analytic prescription of \citet{Bower:2006vw}, hereafter B06.  Galaxies are selected based on their halo mass ($>$10$^{14}$ M$_\odot$ for clusters) and their observed colours and magnitudes in this prescription.  Red-sequence galaxies were selected by fitting a Gaussian to the red peak of the $(B-V$) colour distribution and locating the local minimum on the blue side to isolate the extent of the red-sequence.  In order to capture the uncertainties in sampling the simulation in the same way as the observations, we vary the slope and normalisation of this red-sequence limit (by 15\% and 3\% respectively) as well as the magnitude limits used to define DGR (0.1 mags each limit) and Monte-Carlo the results.  The points for clusters and field in three redshift bins (0.0, 0.5, 1.0) are shown as asterisks and triangles respectively in Fig.~1.  Statistical errors are negligible, so the error bars reflect this systematic variation.

When confronted with the observational data, the semi-analytic model fails to reproduce either the redshift evolution of the DGR or the relative difference between clusters and the field at a given redshift.  The model predicts approximately the correct DGR for clusters in the local universe, but this DGR value then remains constant out to z$\sim$1.  These results are insensitive to the exact halo mass chosen to define the clusters, as can be inferred from the similarity between the cluster and field points at each redshift.  i.e., the DGR is not a strong function of halo mass in the model.

\section{Discussion \& Conclusions}

Previous works have suggested that there appears to be little or no evolution in the bright end of the RSLF, both in clusters and the field (e.g., \citealt{Gilbank:2007rq}; \citealt{Scarlata:2007oz}).  If this is the case, the evolution of the DGR can be attributed to the build-up of faint galaxies on the red-sequence with increasing cosmic time.  This implies that star-formation is terminated in giant galaxies first and later in dwarf galaxies, in agreement with the ``down-sizing'' picture of star-formation.  
Furthermore, the lower value of DGR in the field compared with clusters implies that star-formation was completed in cluster galaxies before field galaxies.  

Performing a similar analysis on the B06 semi-analytic model predicts a non-evolving DGR out to z$=$1, with little difference between clusters and the field.  Such a result suggests that faint galaxies move onto the red-sequence too efficiently in the model, with this build-up of the faint end of the red-sequence already having been completed by the earliest epoch considered here (z$\sim$1).  

The methods available in the models for the quenching of star-formation include: feedback from supernovae or AGN and hot halo gas removal. This latter process only applies to galaxies which are satellites (i.e. not the most massive, central object) in their dark matter halo.  The relative importance of the various quenching mechanisms are summarised in fig.~4 of B06.  At the bright end of the red-sequence, most of the galaxies are the central object in their dark matter halo and thus feedback from AGN or supernova winds are required to turn off star-formation.  At the faint end (in the dwarf regime), red-sequence galaxies are predominantly satellites. We have verified that this is still the case for all the redshifts we consider here.  In most galaxy formation models (e.g., B06, \citealt{De-Lucia:2006sa}), when any galaxy enters a more massive dark matter halo and becomes a satellite, it instantly has all of its hot gas removed in a process generally referred to as `strangulation` \citep{Larson:1980wj,Balogh:2000yv}. Thus, once it has consumed the cold gas already present in its disk, its fuel can no longer be replenished by hot gas from the halo and its star-formation is terminated.  This prescription may be too efficient.  Recent simulations by \citet{McCarthy:2007mr}  found that the removal of the initial hot galactic halo gas does not occur instantaneously, but likely takes $\sim$ 1 Gyr to remove the bulk of it, and that as much as 30\% can still remain after 10 Gyr.  In addition, the models of \cite{Dekel:2006ii}  predict the existence of a critical halo mass ($\sim10^{12}$ M$_\odot$) for haloes, below which strangulation may not occur.  Thus, we expect additional parameters involving strangulation efficiency/timescale and/or a threshold halo mass for strangulation to be important in tuning the models to reproduce the observations presented in Fig.~1.

To illustrate the effect of a threshold halo mass for strangulation, we use the following toy model to predict the evolution of the field DGR.  Field surveys do not just probe isolated (i.e. non-group) galaxies.  Indeed, they are dominated by groups, especially toward lower redshift as cosmic structure grows. Thus it is group-scale physics which is important in determining the properties of galaxies in field surveys.  We assume further that galaxies in haloes below some (group-size) threshold mass, $M_{th}$, have no red-sequence satellites, i.e. DGR $=0$.  We assume that haloes above $M_{th}$ have a DGR given by that of the observed cluster population at the same redshift (from the best-fit cluster line in Fig.~1).  We consider $M_{th} = (10^{12}, 10^{12.5}, 10^{13}$) M$_\odot$ and measure the fractional contribution of haloes in the appropriate mass range from the Millennium Simulation. We overplot the expected evolution of the field DGR for the different threshold masses in Fig.~1.  It can be seen that both the $M_{th} = 10^{12}$ and $10^{12.5}$ M$_\odot$ models give good agreement with the observed field DGR. In order to reproduce the observations so well, our model requires both an increasing contribution to the field by groups due to growth of structure {\it and} an intrinsic evolution of the DGR in haloes above $M_{th}$.  If either of these factors is removed (e.g. if a constant group fraction at all redshifts or a constant cluster DGR at all redshifts is assumed) then the predicted evolution of the field DGR is too flat with respect to the observations.  If groups do not exhibit exactly the same DGR as clusters, but instead are intermediate between clusters and the field (this is currently an open question), then this would effectively shift our predicted lines down and a lower $M_{th}$ would then reproduce the observations.  Of course, a scenario in which the timescale for the removal of hot gas is much longer may fit the observations equally well and this or some alternative combination of threshold mass/timescale cannot be ruled out.  Indeed, the observed evolution of the cluster DGR we have had to include in our model may reflect the effect of a strangulation timescale.

Disentangling the effects of a timescale and threshold mass for the quenching of satellite galaxies is a complex observational problem.  The colour difference between central and satellite red-sequence galaxies may be a useful test of models.  For example, in the B06 prescription, it is predicted that satellite red-sequence galaxies are significantly redder than central red-sequence galaxies.  \citet{van-den-Bosch:2007da} recently measured the average colours of satellite and central galaxies in the local universe from an approximately halo mass-selected sample and found that the former were indeed on average redder.  They also showed that measuring halo masses from statistical samples for $\sim$10$^{12}$ M$_\odot$ systems is not possible with high completeness and thus, if there is a threshold mass for strangulation around this limit, such studies will require detailed targeted observations of groups (e.g., \citealt{Wilman:2005rd}).  An alternative approach to get at the timescale for strangulation is the study of the star formation histories of individual galaxies undergoing transformation (e.g., \citealt{Moran:2007zk}).  It is possible that the observations in Fig.~1, coupled with new strangulation prescriptions in the semi-analytic models may improve constraints on the likely values of the threshold mass and timescale for strangulation. 

\section*{Acknowledgments}

This research was supported by an Early Researcher Award from the
province of Ontario, and by an NSERC Discovery grant.

\label{lastpage}

\begin{thebibliography}{31}
\expandafter\ifx\csname natexlab\endcsname\relax\def\natexlab#1{#1}\fi

\bibitem[{{Andreon}(2007)}]{Andreon:2007jb}
{Andreon}, S. 2007, \mnras~submitted, astro-ph/710.2737

\bibitem[{{Baldry} {et~al.}(2006){Baldry}, {Balogh}, {Bower}, {Glazebrook},
  {Nichol}, {Bamford}, \& {Budavari}}]{Baldry:2006bq}
{Baldry}, I.~K., {Balogh}, M.~L., {Bower}, R.~G., {Glazebrook}, K., {Nichol},
  R.~C., {Bamford}, S.~P., \& {Budavari}, T. 2006, \mnras, 373, 469

\bibitem[{{Baldry} {et~al.}(2004){Baldry}, {Glazebrook}, {Brinkmann},
  {Ivezi{\'c}}, {Lupton}, {Nichol}, \& {Szalay}}]{Baldry:2004wj}
{Baldry}, I.~K., {Glazebrook}, K., {Brinkmann}, J., {Ivezi{\'c}}, {\v Z}.,
  {Lupton}, R.~H., {Nichol}, R.~C., \& {Szalay}, A.~S. 2004, \apj, 600, 681

\bibitem[{{Balogh} {et~al.}(2000){Balogh}, {Navarro}, \&
  {Morris}}]{Balogh:2000yv}
{Balogh}, M.~L., {Navarro}, J.~F., \& {Morris}, S.~L. 2000, \apj, 540, 113

\bibitem[{{Barkhouse} {et~al.}(2007){Barkhouse}, {Yee}, \&
  {Lopez-Cruz}}]{barkhouse07a}
{Barkhouse}, W., {Yee}, H.~K.~C., \& {Lopez-Cruz}, O. 2007, submitted

\bibitem[{{Bell} {et~al.}(2003){Bell}, {McIntosh}, {Katz}, \&
  {Weinberg}}]{Bell:2003iq}
{Bell}, E.~F., {McIntosh}, D.~H., {Katz}, N., \& {Weinberg}, M.~D. 2003, \apjs,
  149, 289

\bibitem[{{Bell} {et~al.}(2004){Bell}, {Wolf}, {Meisenheimer}, {Rix}, {Borch},
  {Dye}, {Kleinheinrich}, {Wisotzki}, \& {McIntosh}}]{Bell:2004lb}
{Bell}, E.~F., {Wolf}, C., {Meisenheimer}, K., {Rix}, H.-W., {Borch}, A.,
  {Dye}, S., {Kleinheinrich}, M., {Wisotzki}, L., \& {McIntosh}, D.~H. 2004,
  \apj, 608, 752

\bibitem[{{Bower} {et~al.}(2006){Bower}, {Benson}, {Malbon}, {Helly}, {Frenk},
  {Baugh}, {Cole}, \& {Lacey}}]{Bower:2006vw}
{Bower}, R.~G., {Benson}, A.~J., {Malbon}, R., {Helly}, J.~C., {Frenk}, C.~S.,
  {Baugh}, C.~M., {Cole}, S., \& {Lacey}, C.~G. 2006, \mnras, 370, 645

\bibitem[{{Brown} {et~al.}(2007){Brown}, {Dey}, {Jannuzi}, {Brand}, {Benson},
  {Brodwin}, {Croton}, \& {Eisenhardt}}]{Brown:2007kk}
{Brown}, M.~J.~I., {Dey}, A., {Jannuzi}, B.~T., {Brand}, K., {Benson}, A.~J.,
  {Brodwin}, M., {Croton}, D.~J., \& {Eisenhardt}, P.~R. 2007, \apj, 654, 858

\bibitem[{{Bruzual} \& {Charlot}(2003)}]{Bruzual:2003de}
{Bruzual}, G. \& {Charlot}, S. 2003, \mnras, 344, 1000

\bibitem[{{Carlberg} {et~al.}(1997){Carlberg}, {Yee}, {Ellingson}, {Morris},
  {Abraham}, {Gravel}, {Pritchet}, {Smecker-Hane}, {Hartwick}, {Hesser},
  {Hutchings}, \& {Oke}}]{carlberg97}
{Carlberg}, R.~G., et al.\ 1997, \apjl, 485, L13

\bibitem[{{Cimatti} {et~al.}(2006){Cimatti}, {Daddi}, \&
  {Renzini}}]{Cimatti:2006vz}
{Cimatti}, A., {Daddi}, E., \& {Renzini}, A. 2006, \aap, 453, L29

\bibitem[{{Cowie} {et~al.}(1996){Cowie}, {Songaila}, {Hu}, \&
  {Cohen}}]{Cowie:1996xw}
{Cowie}, L.~L., {Songaila}, A., {Hu}, E.~M., \& {Cohen}, J.~G. 1996, \aj, 112,
  839

\bibitem[{{De Lucia} {et~al.}(2004){De Lucia}, {Poggianti},
  {Arag{\'o}n-Salamanca}, {Clowe}, {Halliday}, {Jablonka}, {Milvang-Jensen},
  {Pell{\'o}}, {Poirier}, {Rudnick}, {Saglia}, {Simard}, \&
  {White}}]{De-Lucia:2004xa}
{De Lucia}, et al.\ 2004, \apjl, 610, L77

\bibitem[{{De Lucia} {et~al.}(2006){De Lucia}, {Springel}, {White}, {Croton},
  \& {Kauffmann}}]{De-Lucia:2006sa}
{De Lucia}, G., {Springel}, V., {White}, S.~D.~M., {Croton}, D., \&
  {Kauffmann}, G. 2006, \mnras, 366, 499

\bibitem[{{de Lucia} {et~al.}(2007){de Lucia}, {Poggianti},
  {Arag{\'o}n-Salamanca}, {White}, {Zaritsky}, {Clowe}, {Halliday}, {Jablonka},
  {von der Linden}, {Milvang-Jensen}, {Pell{\'o}}, {Rudnick}, {Saglia}, \&
  {Simard}}]{de-Lucia:2007li}
{de Lucia}, et a;.\  2007, \mnras, 374, 809

\bibitem[{{Dekel} \& {Birnboim}(2006)}]{Dekel:2006ii}
{Dekel}, A. \& {Birnboim}, Y. 2006, \mnras, 368, 2

\bibitem[{{Driver} {et~al.}(2006){Driver}, {Allen}, {Graham}, {Cameron},
  {Liske}, {Ellis}, {Cross}, {De Propris}, {Phillipps}, \&
  {Couch}}]{Driver:2006li}
{Driver}, S.~P., et al.\ 2006, \mnras, 368, 414

\bibitem[{{Gilbank} {et~al.}(2007){Gilbank}, {Yee}, {Ellingson}, {Gladders},
  {Loh}, {Barrientos}, \& {Barkhouse}}]{Gilbank:2007rq}
{Gilbank}, D.~G., {Yee}, H.~K.~C., {Ellingson}, E., {Gladders}, M.~D., {Loh},
  Y.~., {Barrientos}, L.~F., \& {Barkhouse}, W.~A. 2007, \apj~accepted, astro-ph/710.2351

\bibitem[{{Hansen} {et~al.}(2007){Hansen}, {Sheldon}, {Wechsler}, \&
  {Koester}}]{Hansen:2007bb}
{Hansen}, S.~M., {Sheldon}, E.~S., {Wechsler}, R.~H., \& {Koester}, B.~P. 2007,
  \apj~submitted, astro-ph/710.3780

\bibitem[{{Larson} {et~al.}(1980){Larson}, {Tinsley}, \&
  {Caldwell}}]{Larson:1980wj}
{Larson}, R.~B., {Tinsley}, B.~M., \& {Caldwell}, C.~N. 1980, \apj, 237, 692

\bibitem[{{McCarthy} {et~al.}(2007){McCarthy}, {Frenk}, {Font}, {Lacey},
  {Bower}, {Mitchell}, {Balogh}, \& {Theuns}}]{McCarthy:2007mr}
{McCarthy}, I.~G., {Frenk}, C.~S., {Font}, A.~S., {Lacey}, C.~G., {Bower},
  R.~G., {Mitchell}, N.~L., {Balogh}, M.~L., \& {Theuns}, T. 2007, \mnras~accepted, astro-ph/710.0964

\bibitem[{{Miller} {et~al.}(2005){Miller}, {Nichol}, {Reichart}, {Wechsler},
  {Evrard}, {Annis}, {McKay}, {Bahcall}, {Bernardi}, {Boehringer}, {Connolly},
  {Goto}, {Kniazev}, {Lamb}, {Postman}, {Schneider}, {Sheth}, \&
  {Voges}}]{Miller:2005qb}
{Miller}, C.~J., et al.\ 2005, \aj,
  130, 968

\bibitem[{{Moran} {et~al.}(2007){Moran}, {Loh}, {Ellis}, {Treu}, {Bundy}, \&
  {MacArthur}}]{Moran:2007zk}
{Moran}, S.~M., {Loh}, B.~L., {Ellis}, R.~S., {Treu}, T., {Bundy}, K., \&
  {MacArthur}, L.~A. 2007, \apj, 665, 1067

\bibitem[{{Scarlata} {et~al.}(2007){Scarlata}, {Carollo}, {Lilly}, {Feldmann},
  {Kampczyk}, {Renzini}, {Cimatti}, {Halliday}, {Daddi}, {Sargent},
  {Koekemoer}, {Scoville}, {Kneib}, {Leauthaud}, {Massey}, {Rhodes}, {Tasca},
  {Capak}, {McCracken}, {Mobasher}, {Taniguchi}, {Thompson}, {Ajiki}, {Aussel},
  {Murayama}, {Sanders}, {Sasaki}, {Shioya}, \& {Takahashi}}]{Scarlata:2007oz}
{Scarlata}, C., et al.\ 2007, \apjs, 172, 494

\bibitem[{{Stott} {et~al.}(2007){Stott}, {Smail}, {Edge}, {Ebeling}, {Smith},
  {Kneib}, \& {Pimbblet}}]{Stott:2007wc}
{Stott}, J.~P., {Smail}, I., {Edge}, A.~C., {Ebeling}, H., {Smith}, G.~P.,
  {Kneib}, J.~., \& {Pimbblet}, K.~A. 2007, \apj, 661, 95

\bibitem[{{Tanaka} {et~al.}(2005){Tanaka}, {Kodama}, {Arimoto}, {Okamura},
  {Umetsu}, {Shimasaku}, {Tanaka}, \& {Yamada}}]{Tanaka:2005mk}
{Tanaka}, M., {Kodama}, T., {Arimoto}, N., {Okamura}, S., {Umetsu}, K.,
  {Shimasaku}, K., {Tanaka}, I., \& {Yamada}, T. 2005, \mnras, 362, 268

\bibitem[{{van den Bosch} {et~al.}(2007){van den Bosch}, {Aquino}, {Yang},
  {Mo}, {Pasquali}, {McIntosh}, {Weinmann}, \& {Kang}}]{van-den-Bosch:2007da}
{van den Bosch}, F.~C., {Aquino}, D., {Yang}, X., {Mo}, H.~J., {Pasquali}, A.,
  {McIntosh}, D.~H., {Weinmann}, S.~M., \& {Kang}, X. 2007, ArXiv e-prints, 710

\bibitem[{{Weinmann} {et~al.}(2006){Weinmann}, {van den Bosch}, {Yang}, {Mo},
  {Croton}, \& {Moore}}]{Weinmann:2006cn}
{Weinmann}, S.~M., {van den Bosch}, F.~C., {Yang}, X., {Mo}, H.~J., {Croton},
  D.~J., \& {Moore}, B. 2006, \mnras, 372, 1161

\bibitem[{{Wilman} {et~al.}(2005){Wilman}, {Balogh}, {Bower}, {Mulchaey},
  {Oemler}, {Carlberg}, {Morris}, \& {Whitaker}}]{Wilman:2005rd}
{Wilman}, D.~J., {Balogh}, M.~L., {Bower}, R.~G., {Mulchaey}, J.~S., {Oemler},
  A., {Carlberg}, R.~G., {Morris}, S.~L., \& {Whitaker}, R.~J. 2005, \mnras,
  358, 71

\bibitem[{{Zucca} {et~al.}(2006){Zucca}, {Ilbert}, {Bardelli}, {Tresse},
  {Zamorani}, {Arnouts}, {Pozzetti}, {Bolzonella}, {McCracken}, {Bottini},
  {Garilli}, {Le Brun}, {Le F{\`e}vre}, {Maccagni}, {Picat}, {Scaramella},
  {Scodeggio}, {Vettolani}, {Zanichelli}, {Adami}, {Arnaboldi}, {Cappi},
  {Charlot}, {Ciliegi}, {Contini}, {Foucaud}, {Franzetti}, {Gavignaud},
  {Guzzo}, {Iovino}, {Marano}, {Marinoni}, {Mazure}, {Meneux}, {Merighi},
  {Paltani}, {Pell{\`o}}, {Pollo}, {Radovich}, {Bondi}, {Bongiorno},
  {Busarello}, {Cucciati}, {Gregorini}, {Lamareille}, {Mathez}, {Mellier},
  {Merluzzi}, {Ripepi}, \& {Rizzo}}]{Zucca:2006oz}
{Zucca}, E., et al.\ 2006, \aap, 455, 879

\end{thebibliography}
\end{document}